\documentclass{aa501}
\usepackage{graphics,amssymb,natbib,rotating,psfrag,amsmath,epsfig,array}

\title{Cumulative light curves of gamma-ray bursts and relaxation systems}
\titlerunning{Cumulative light curves of GRBs}

\author{S.\,McBreen\inst{1} \and
        B.\,McBreen\inst{1} \and
        L.\,Hanlon\inst{1} \and
        F.\,Quilligan\inst{1,2}}

\institute{Department of Experimental Physics, University College
Dublin, Dublin 4, Ireland \and Intel Corporation, Leixlip, Co.
Kildare, Ireland}

\offprints{smcbreen@bermuda.ucd.ie}
\date{Received / Accepted}

\abstract{The cumulative light curves of a large sample of gamma
ray bursts (GRBs) were obtained by summing the BATSE counts. The
smoothed profiles are much simpler than the complex and erratic
running light curves that are normally used.  For most GRBs the
slope of the cumulative light curve (S) is approximately constant
over a large fraction of the burst.  The bursts are modelled as
relaxation systems that continuously accumulate energy in the
reservoir and discontinuously release it.  The slope is a measure
of the cumulative power output of the central engine. A plot of S
versus peak flux in 64 ms (P$_{64 {\rm ms}}$) shows a very good
correlation over a wide range for both long and short GRBs. No
relationship was found between S and the GRBs with known
redshift.  The standard slope (S$^{\prime}$), which is
representative of the power output per unit time, is correlated
separately with P$_{64 {\rm ms}}$ for both sub-classes indicating
more powerful outbursts for the short GRBs.  S$^{\prime}$ is
also anticorrelated with GRB duration. These results imply that GRBs are
powered by accretion into a black hole.
\keywords{Gamma rays -- bursts: Gamma rays -- observations:
Methods -- data analysis: Methods -- statistical}
}

\begin{document}
 \maketitle

\section{Introduction}

Cosmological gamma ray bursts (GRBs) emit an extraordinary amount
of energy in gamma rays \citep{cfpc:1997,vanpara:1997,piran:1999}.
The source of this energy may be a cataclysmic event involving
mergers of compact objects such as neutron star binaries or
neutron stars and black hole binaries \citep{ruffjan:1999} or the
formation of a black hole during or after the collapse of massive
stars \citep{reemes:1994,pacy:1998,vietri:1998,macfad:1999,reeves:2002}. GRB
light curves are complex and erratic \citep{fishman:1995}.  There
are a number of recent results on the properties of the pulses in
GRBs and their relationship to the duration T$_{90}$ (Ramirez-Ruiz \&
Fenimore, 2000; Salmonson, 2000; Norris, 2002; McBreen et al., 2002a; Gupta et al.,
2002; Quilligan et al., 2002 and references therein). The correlated
pulses in GRBs have a unique set of properties. In this
paper we show in sections 3 and 4 that the slope of the
cumulative light curve of most short and long GRBs is
approximately linear and the bursts can be modelled as relaxation
systems.  In addition the slopes are highly correlated with the
peak flux and anticorrelated with GRB duration. The large and
uniform BATSE sample of GRBs were used in this analysis
\citep{fishman:1995,kmf:1993}.

\section{Timing analysis of the GRB profiles}

A large sample of 498 of the brightest BATSE GRBs with data combined from
the four energy channels was analysed and denoised using either
wavelets or median filters. A detailed account of this process
including sample selection and pulse analysis in GRBs has been
given elsewhere \citep{quilligan:2002,sheila:2001}.  The GRB
sample was restricted here to bursts that had the on-board summed
count from only two Large Area Detectors and includes 55 GRBs
with T$_{90} <$ 2 s analysed at 5 ms resolution, 250 GRBs with
T$_{90}
>$ 2 s at 64 ms resolution.  To extend the sample to include additional GRBs
with T$_{90} >$ 100 s, a further 71 GRBs were included and
analysed at 256 ms resolution.  The cumulative light curve was
obtained by converting the BATSE rates to counts and taking the
cumulative sum.  The cumulative count from the same number of on-board
detectors is uncorrected for the
detector response and the angles between the detectors and GRB
sources.  The errors from these effects should be less than $\pm
30\%$.  As a check on this procedure, the cumulative count for
the GRBs was compared with the BATSE fluences. There is a high
degree of correlation between the two quantities showing that the
cumulative count from the same number of detectors is a good
measure of the burst.  There is close agreement with the
cumulative light curves obtained in the spectroscopic sample of
\citet{pebs:2000}.

\section{Results}

The running and cumulative light curves of a sample of bursts,
that cover a wide range in T$_{90}$, are given in Fig. 1. The
cumulative light curves smooth the running profiles. The slope of
the cumulative light curve (S) was measured for the section of
the GRB that included more than 50$\%$ of the total counts and
could be represented by a straight line, with a precision in the
measured slope that was always much better than a factor of 2.
Not all GRBs were fit with a single straight line because of long
time intervals with very little or no emission (Fig. 1d) and some
of these GRBs have been discussed by \citet{rm:2001}.  In these
cases two or more straight lines were used to fit the separate
periods of emission.  The average value of the slopes was adopted
for that GRB.  This procedure was used for 66 GRBs with T$_{90}
>$ 2 s and 2 GRBs with T$_{90} <$ 2 s.  The cumulative light
curves of a number of GRBs have measurable curvature and in these
cases a curve is a more accurate fit than the straight line
adopted here. These GRBs are the subject of a separate
publication \citep{mcbreenb:2002}.

The median percentage of the integrated counts over which the
slope was measured is 71\% for GRBs with T$_{90} >$ 2 s and 83\%
for short GRBs.  The corresponding values of the median
percentages of T$_{90}$/T$_{50}$ are 34\%/127\% and 68\%/140\%.
T$_{50}$ is a better measure than T$_{90}$ of the active period of
the burst over which the slopes were measured.  It is also better
anticorrelated with the slope than T$_{90}$ (Table 1, Fig. 2c).
This effect is illustrated in Fig. 1d where 75\% of the
integrated counts is used for the slope over only 24\% of
T$_{90}$ whereas in Fig. 1a the corresponding values are 69\% and
59\% respectively.  There is a long quiescent interval in the GRB
in Fig. 1d which contributes to the small percentage value of
T$_{90}$.

The slopes of the cumulative light curves of 376 GRBs are plotted
in Fig. 2a versus the peak flux in 64 ms (P$_{64 \rm ms}$). There
is a good correlation between S and P$_{64 \rm ms}$. Spearman
rank order correlation coefficients $\rho$, associated
probabilities and index of best fit power law for these quantities
are listed in Table 1. The standard slope S$^{\prime}$ was
obtained by dividing the slope of the cumulative counts (S) by
the time over which the slope was measured and hence it
represents the cumulative counts or power output of the source
per unit time. The standard slope is plotted versus P$_{64 \rm
ms}$ in Fig. 2b. The correlation coefficients are listed in Table
1 for short and long GRBs.  The best fit power laws have indices
close to 1.6 for  both classes of GRBs but there is considerable
spread in the values with the short bursts displaced from the
long bursts by $\sim 10^{2}$.

The standard slope is plotted versus T$_{50}$ in Fig. 2c.  The
correlation coefficient is -0.90 with a best fit power law index
of -1.1.  The corresponding values for T$_{90}$ are -0.85 and
-1.08 respectively.  The values are also listed in Table 1 for
T$_{90}$ and T$_{50}$ versus S.  S$^{\prime}$ is better
anticorrelated with duration than S.

\citet{fenimore:1999} obtained the average temporal profile of 98
GRBs by normalising to a standard duration and a standard peak
counts.  The average profile is reasonably flat over more than
50\% of the duration (Fig. 1b in Fenimore (1999)) implying a
linear increase in the cumulative profile and hence in agreement
with the results presented here.

\small
		\begin{table} [t]
 \caption{Spearman rank order correlation coefficients, $\rho$, the associated
 probabilities and index of the best fit power law for a range of GRB properties. In all cases the high values of $\rho$
show either strong correlations or anticorrelations between the
burst properties.}
\begin{tabular}{lccc}\hline\hline 
\raisebox{-1.5ex}[0cm][0cm]{Properties} &\raisebox{-1.5ex}[0cm][0cm]{
$\rho$} & \raisebox{-1.5ex}[0cm][0cm]{Probability }& Power Law \\
    &   & &  Index\\
\hline
P$_{64 \rm ms}$ vs. S & \,\,0.84 & $<10^{-48}$ & 1.04 \\
P$_{64 \rm ms}$ vs. S$^{\prime}$ (T$_{90}>2\,s$) & \,\,0.72 & $<10^{-48}$ & 1.63 \\
P$_{64 \rm ms}$ vs. S$^{\prime}$ (T$_{90}<2\,s$) & \,\,0.69 & $6.1
\times 10^{-9}$  & 1.65 \\
T$_{90}$ vs. S$^{\prime}$ & $-$0.85 & $<10^{-48}$ & $-$1.08 \\
T$_{90}$ vs. S (T$_{90}>2$\,s)& $-$0.58 & $1.7 \times 10^{-30}$ & $-0.54$  \\
T$_{90}$ vs. S (T$_{90}<2$\,s)& $-$0.39 & $3.0 \times 10^{-3}$ & $-$0.37 \\
T$_{50}$ vs. S$^{\prime}$ & $-$0.9 & $<10^{-48}$ & $-$1.10 \\
T$_{50}$ vs. S  (T$_{90}>2$\,s)& $-$0.64 & $1.9 \times 10^{-31}$ & $-$0.53 \\
T$_{50}$ vs. S  (T$_{90}<2$\,s)& $-$0.53 & $3.7 \times 10^{-5}$ & $-$0.32\\
Fluence vs.  & \raisebox{-1.5ex}[0cm][0cm]{\,\, 0.92} &
\raisebox{-1.5ex}[0cm][0cm]{$<10^{-48}$} &
\raisebox{-1.5ex}[0cm][0cm]{\,\,\,\,0.92} \\
\quad Cumulative counts\\
\hline
\end{tabular}
\end{table}
\normalsize

\begin{figure}[t]
 \leavevmode
 \psfrag{Time}[t]{\Large Time (sec)}
 \psfrag{Relative Time}[t]{\large Time (sec)}
 \psfrag{Normalised Fit}[t]{\large }
 \psfrag{Counts}[t]{\Large Counts}
 \psfrag{Integrated Counts}[tc]{\Large Cumulative Counts}
\begin{center}
\resizebox{0.9\columnwidth}{0.14\textheight}{\includegraphics{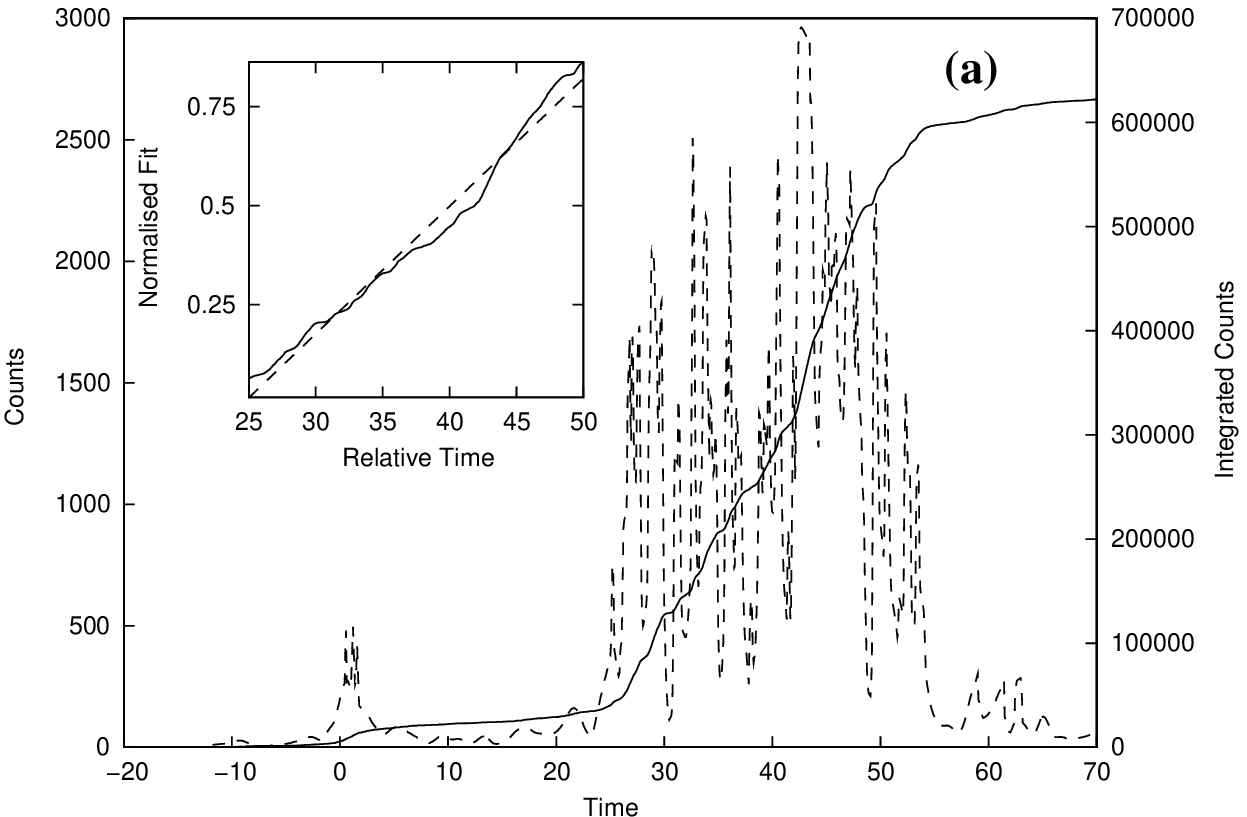}}\\[8.0pt]
\resizebox{0.9\columnwidth}{0.14\textheight}{\includegraphics{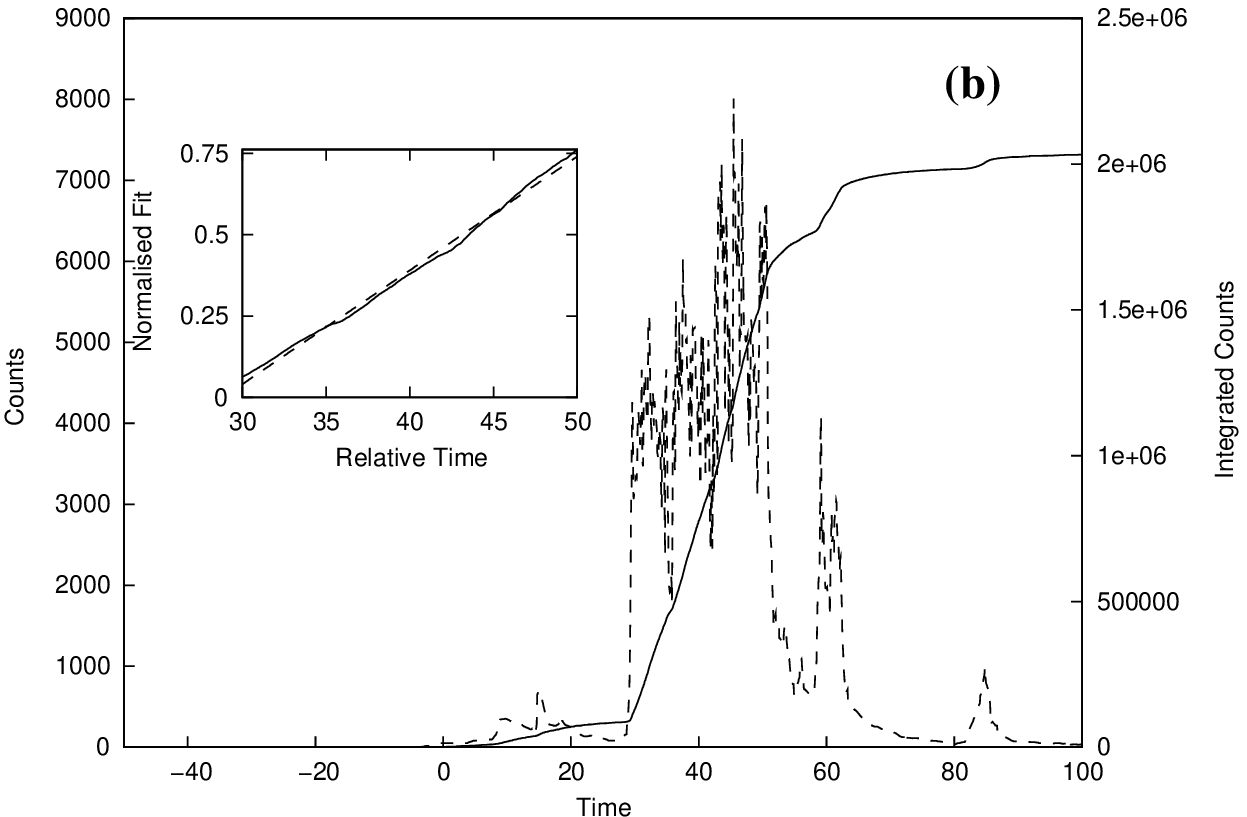}}\\[8.0pt]
\resizebox{0.9\columnwidth}{0.14\textheight}{\includegraphics{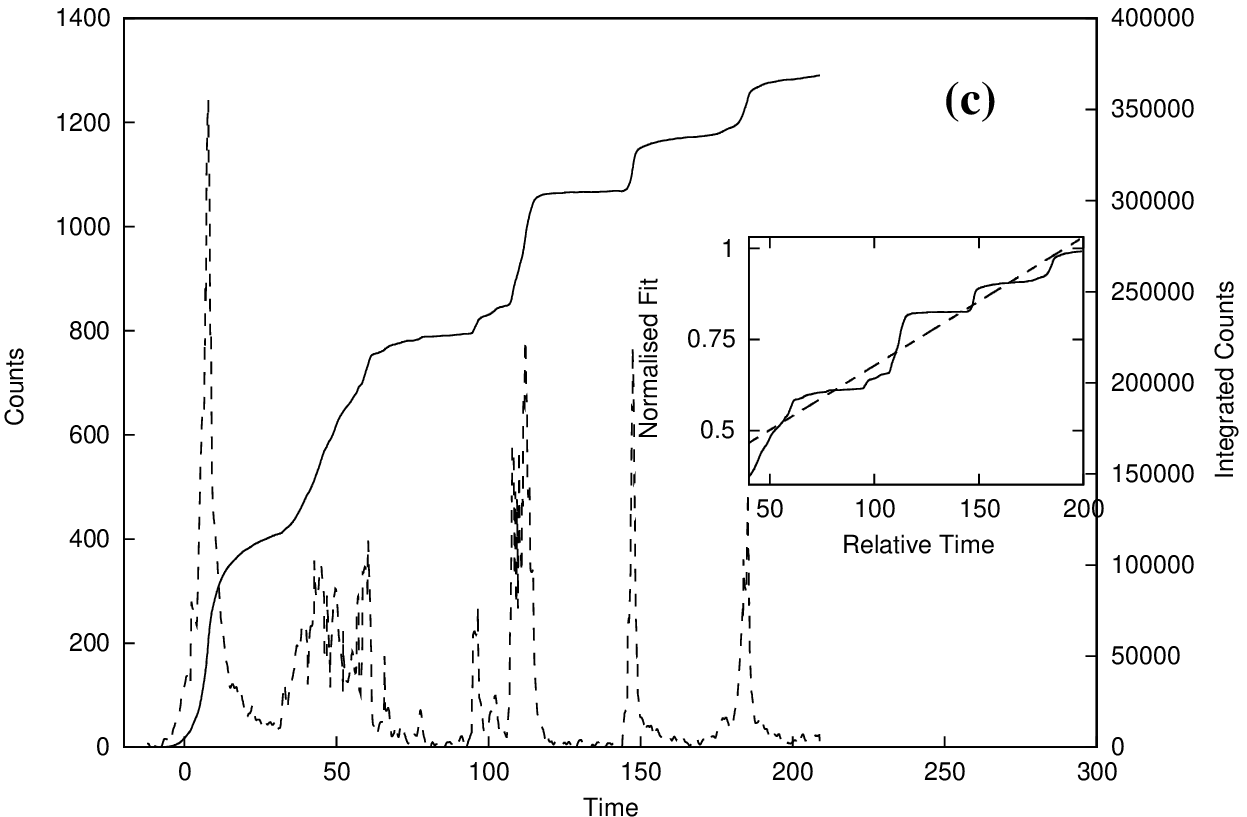}}\\[8.0pt]
\resizebox{0.9\columnwidth}{0.14\textheight}{\includegraphics{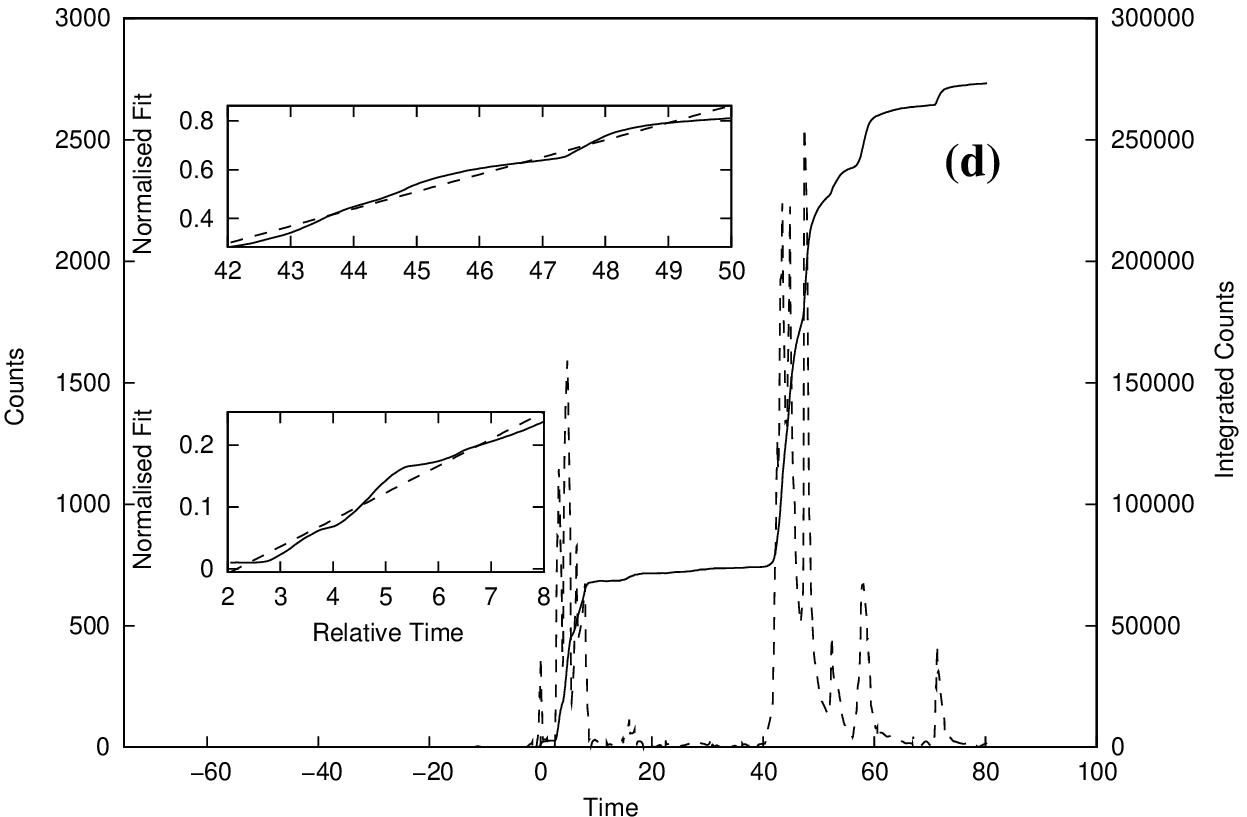}}\\[8.0pt]
\resizebox{0.9\columnwidth}{0.14\textheight}{\includegraphics{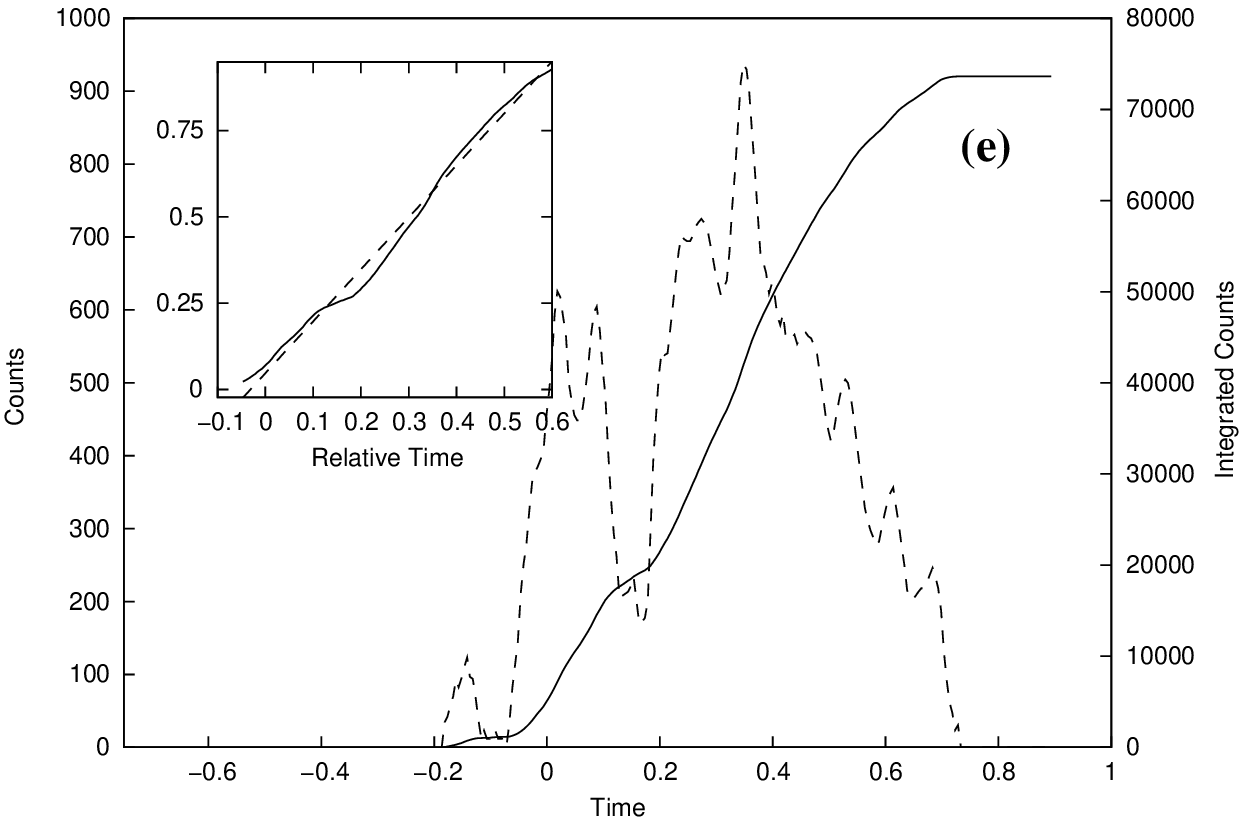}}
    \caption{The running (dashed) and cumulative (solid) light curves of the BATSE bursts with trigger
    numbers a) 3128, b) 3057, c) 3042, d) 7560 , e) 2217 with
    count per 64 ms and cumulative count scales on the left and right vertical axes.  The insert
    gives the straight line fit (dashed) to the
    cumulative count (solid) for the relevant section(s) of the
    GRB. The vertical axes in the inserts are the normalised cumulative count.}
     \end{center}
\end{figure}

\section{Discussion}
\subsection{GRBs as relaxation systems}

It can be assumed that the sum of the counts in the bursts (Fig.
1.) is a good measure of the integrated energy emitted by the
source because the peak energy lies well within the BATSE band
\citep{fishman:1995}.  Many models of GRBs consist of a newly
formed black hole that accretes from a remnant torus that is
cooled by neutrino emission
\citep{popham:1999,narayan:2001,leeram:2002}. The energy to drive
the relativistic jets and bursts may be extracted from the disk
and spinning black hole by MHD processes and neutrino
annihilation. The efficiency of these processes in creating
relativistic jets is on the order of one percent
\citep{macfad:1999}

Models of this type can be usefully compared with a relaxation
system \citep{palmer:1999} which is taken to be one that
continuously accumulates energy from the accretion process and
discontinuously releases it. The energy in the reservoir at any
time t is
\begin{equation}
E(t) = E_{o} + \int_{o}^{t} R(t)dt - \Sigma S_{i}
\end{equation}
where E$_{o}$ is the energy stored in the reservoir that
accumulates energy at a rate R(t) and discontinuously releases
events of size S$_{i}$.

The simplest system is referred to as a relaxation oscillator
where there is a fixed level or trip-point that triggers a
release of the energy when E = E$_{\rm max}$.  The soft gamma
ray repeater (SGR)
\citep{palmer:1999,gwk:2000,hmrs:1994} and GRB pulses are not
consistent with this oscillator.  More complicated behaviour
occurs when the accumulation rate, trigger rate or release
strength are not constant.  If the system starts from a minimum
level E = E$_{\rm min}$, accumulates energy at a constant rate R
= r, the sum of the releases is approximately a linear function
of time i.e. \(\Sigma S_{i} \propto rt\).  This model can account
for the approximately linear increase in cumulative counts from
GRBs (Fig 1).  The pulses in GRBs have a tendency to keep the cumulative
count close to a linear function and maintain a steady state situation.

\citet{rm:2001} found a correlation between the duration of an
emission episode in a multi-peaked burst and the duration of the
preceding quiescent time which is similar to the above scenario.
The system could build up its energy probably via an MHD
instability driven dynamo and reach a near critical or metastable
level.  A local instability could cause a rapid dissipation of
all the stored energy.  The system will tend to return to a more
stable configuration characterised by a certain threshold energy
E$_{\rm o}$, or a sub-critical magnetic field configuration.  The
source then becomes quiescent.  Interestingly both models can be
unified by noting that each time the system is completely drained
by a total release of accumulated energy, the central engine goes
quiescent but otherwise the energy extraction is usually in
episodes that are incomplete releases of energy. In this case the
longer the quiescent time, the higher the stored energy from the
next episode.  Such a situation may give rise to the observed
correlations between long quiescent times \citep{rm:2001}, correlated pulse
properties and intervals between pulses
\citep{quilligan:2002,smcb:2002,nakar:2002}.  This is a different
mechanism from any relaxation oscillator which forces a release
of energy when the system reaches an upper level.

\begin{figure}[ht]
\leavevmode

\begin{center}
 \psfrag{Slope}[t]{\large Slope (Cumulative Counts s$^{-1}$) }
 \psfrag{P64ms}[t]{\large P$_{\rm 64 ms} ($ photons cm$^{-2}$ s$^{-1}$)} \
\resizebox{0.9\columnwidth}{0.22\textheight}{\includegraphics{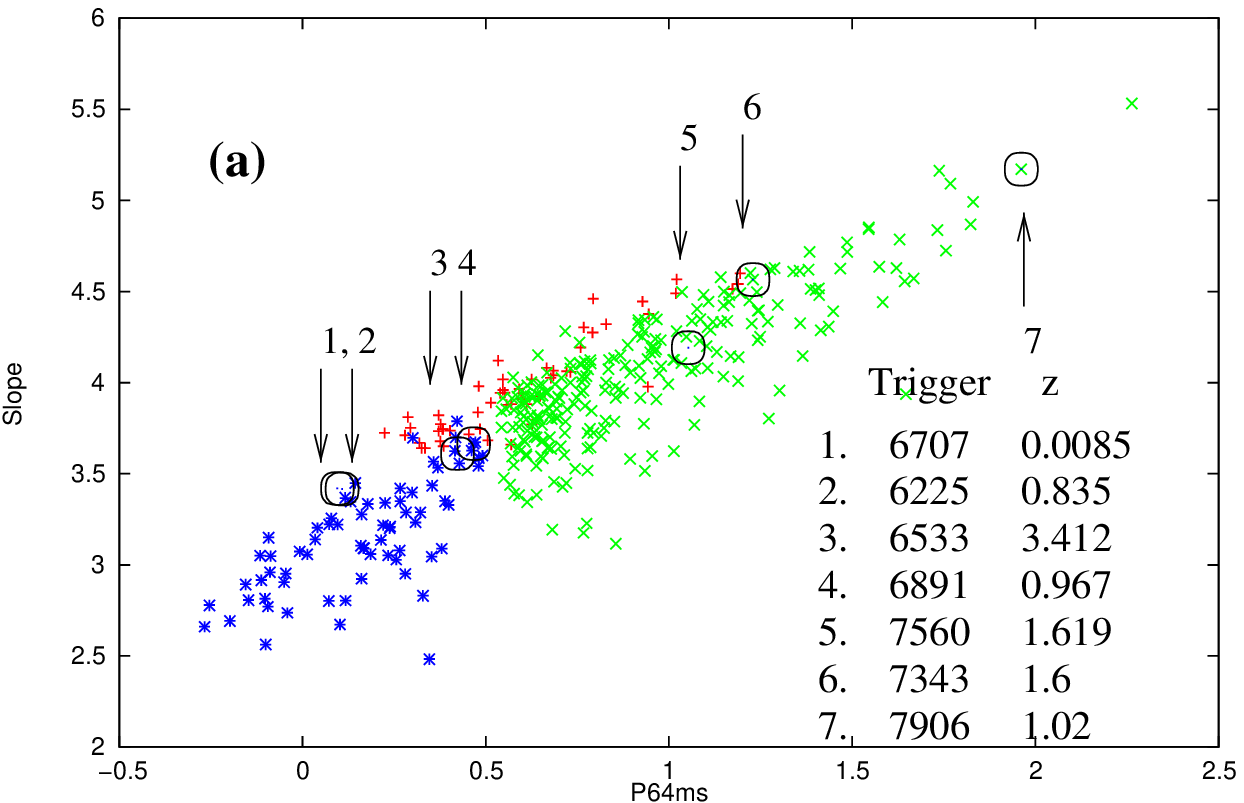}}\\[7.5pt]

 \psfrag{Sprime}[t]{\large Standard Slope (Cumulative Counts s$^{-2}$) }
 \resizebox{0.9\columnwidth}{0.22\textheight}{\includegraphics{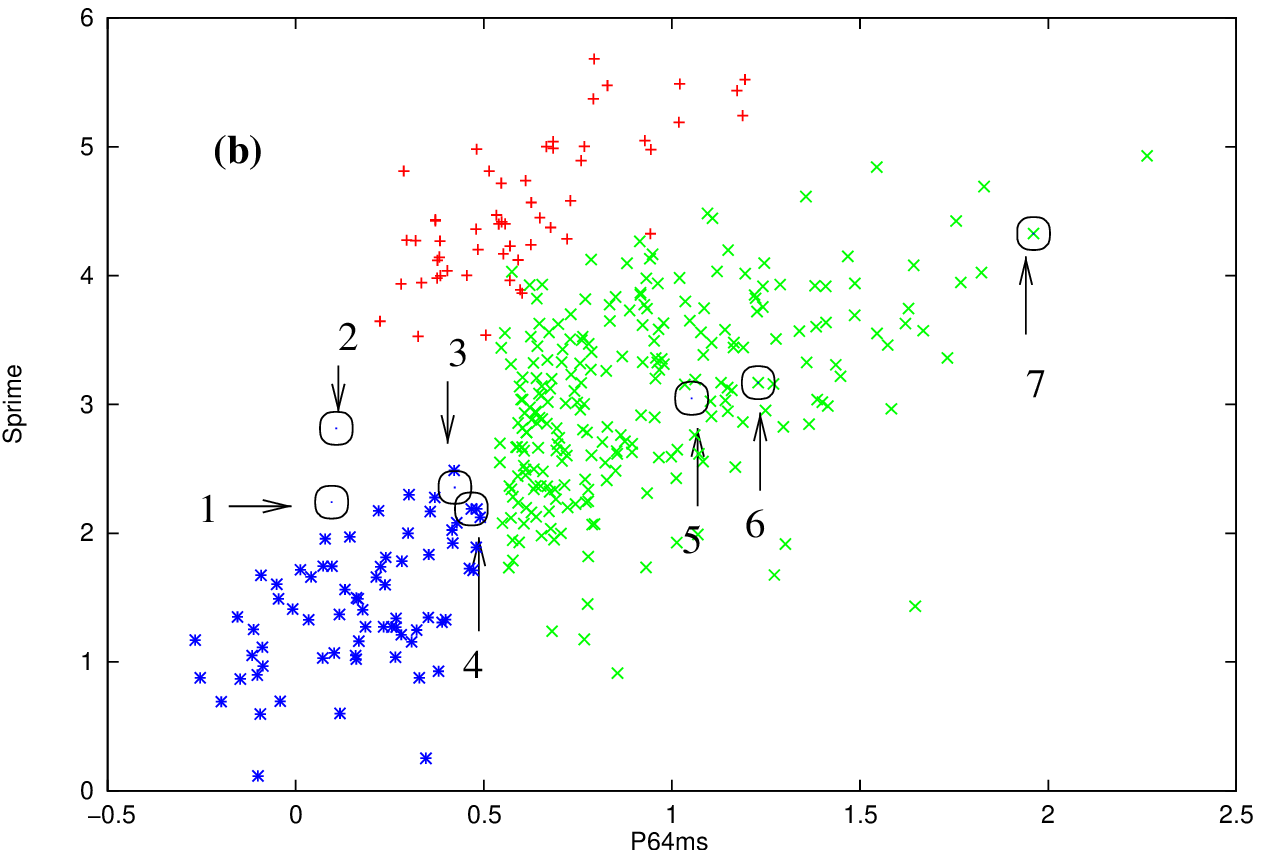}}\\[7.5pt]

 \psfrag{T50}[t]{\large T$_{50}$ (sec) }
 \psfrag{Sprime}[t]{\large Standard Slope (Cumulative Counts s$^{-2}$) }
 \resizebox{0.9\columnwidth}{0.22\textheight}{\includegraphics{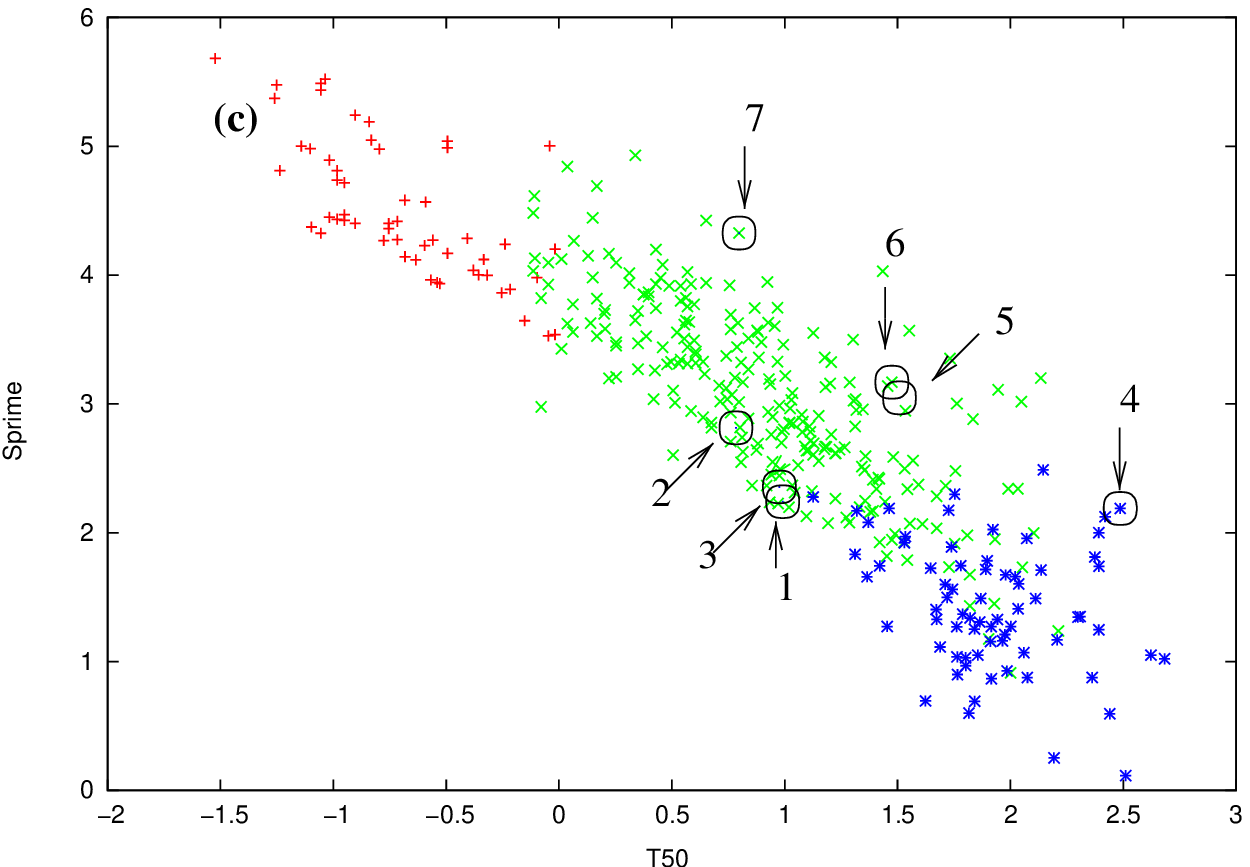}}

    \caption{ The values of P$_{64 \rm ms}$ are plotted versus a) the slope S and b) the standard
    slope S$^{\prime}$ of the GRB cumulative light curve for three categories of GRBs
    i.e. T$_{90} <$ 2 s (red), T$_{90} >$ 2 s (green) and the additional sample with T$_{90} >$ 100 s
    (blue).  T$_{50}$ is plotted versus S$^{\prime}$ in c) for the same three catagories.
     The seven GRBs with known
     redshift and detected by BATSE are labelled (van Paradijs et al., 2000; Castro-Tirado, 2001 and references therein).
     The BATSE trigger numbers and redshifts are given in the top figure.  An extension of the peak flux limited
     sample with T$_{90} >$ 2 s to lower values should populate
     the region containing GRBs 1 and 2 with known z in b).}
 \end{center}
\end{figure}

\subsection{Relationships between the slopes and peak flux}

There is a significant correlation, that extends over a range of
$\sim10^{3}$, between P$_{64 \rm ms}$ and S (Fig. 2a and Table
1).  The seven GRBs, detected by BATSE, with measured redshifts
are also plotted in Fig. 2.  These sources cover a wide range with
no obvious relationship between S, P$_{\rm 64 ms}$ and z.

The separation of GRBs into two sub-classes was characterised by
durations $>$ 2 s and $<$ 2 s \citep{kmf:1993}. The plot of the
standard slope versus P$_{\rm 64 ms}$ (Fig. 2b) also separates the
GRBs into two classes with the short GRBs having a more powerful
output power per unit time.  The cosmological GRBs with known redshift have
values of S and S$^{\prime}$ that range from $7 \times 10^{44}$ to  
$5 \times 10^{47}$ Watts and $1.5 \times 10^{44}$ to $2 \times 10^{46}$ Watts
per second respectively.
There is no evidence for a new class
of GRBs with a different relationship between the standard slope
and P$_{\rm 64 ms}$.

\subsection{Relationships between the slopes and durations}

The data presented in Fig. 2c shows that as the standard slope
increases the values of T$_{50}$ decrease and this effect is
present in both sub-classes (Table 1).  The trend is quite
revealing and shows that the smaller the value of T$_{50}$ the
greater the standard slope or the cumulative power output per
second from the source.  The standard slope plays an important
role in determining T$_{50}$.  
The median values of the pulse properties and time intervals
between pulses were found to increase with T$_{90}$ \citep{smcb:2002}.
The opposite
relationship exists here between S$^{\prime}$ and burst duration
(Table 1) implying that as the standard slope increases there is
a corresponding decrease in the pulse properties. GRBs with high
accretion rates have large values of the standard slope, fast
pulses and short durations whereas lower accretion gives lower
values of S$^{\prime}$, slower pulses that are further apart and
larger values of T$_{90}$/T$_{50}$. These results provide strong
evidence that GRBs are powered by hyperaccretion into a black hole
from a standard type engine
\citep{sal:2000,frail:2001,panait:2001,piran:2001}.

\small
\bibliography{katmonic,allrefs_ch}

\begin{thebibliography}{30}
\expandafter\ifx\csname natexlab\endcsname\relax\def\natexlab#1{#1}\fi

\bibitem[{Castro-Tirado (2001)}]{alberto:2001}
Castro-Tirado, A.~J. 2001, [astro-ph/0102122]

\bibitem[{{Costa} {et~al.}(1997){Costa}, {Frontera}, {Heise}, \& {et al}}]{cfpc:1997}
{Costa}, E., {Frontera}, F., {Heise}, J. et~al. 1997, Nature, 387, 783
  
\bibitem[{{Fenimore}(1999)}]{fenimore:1999}
{Fenimore}, E.~E. 1999, ApJ, 518, 375  
  
\bibitem[{{Fishman} \& {Meegan}(1995)}]{fishman:1995}
{Fishman}, G.~J. \& {Meegan}, C.~A. 1995, Ann. Rev. Astron. Ap., 33, 415

\bibitem[{Frail {et~al.}(2001)Frail, Kulkarni, Sari, {et al}}]{frail:2001}
Frail, D.~A., Kulkarni, S.~R., Sari, R., {et al}. 2001, ApJ, 562, L55

\bibitem[{{G\"{o}g\"{u}s} {et~al.}(2000){G\"{o}g\"{u}s}, {Woods},
  {Kouveliotou}, \& {et al}}]{gwk:2000}
{G\"{o}g\"{u}s}, E., {Woods}, P.~M., {Kouveliotou}, C. {et al}. 2000, ApJ, 532, L121

\bibitem[{{Gupta} {et~al.}(2002){Gupta}, {Das Gupta}, \& {Bhat}}]{gupta:2002}
{Gupta}, V., {Das Gupta}, P. \& {Bhat}, P. 2002, [astro-ph/0206402] 

\bibitem[{{Hurley} {et~al.}(1994){Hurley}, {McBreen}, {Rabbette}, \&
  {Steel}}]{hmrs:1994}
{Hurley}, K.~J., {McBreen}, B., {Rabbette}, M., \& {Steel}, S. 1994, A\&A, 288,
  L49

\bibitem[{{Kouveliotou} {et~al.}(1993){Kouveliotou}, {Meegan}, {Fishman},
  et~al}]{kmf:1993}
{Kouveliotou}, C., {Meegan}, C.~A., {Fishman}, G.~J., et~al. 1993, ApJ, 413,
  L101

\bibitem[{{Lee} \& {Ramirez-Ruiz}(2002)}]{leeram:2002}
{Lee}, W. H. \& {Ramirez-Ruiz}, E. 2002, [astro-ph/0206011]
 
\bibitem[{{MacFadyen} \& {Woosley}(1999)}]{macfad:1999}
{MacFadyen}, A.~I. \& {Woosley}, S.~E. 1999, ApJ, 524, 262

\bibitem[{{McBreen} {et~al.}(2002a){McBreen}, {McBreen}, {Quilligan}, \&
  {Hanlon}}]{smcb:2002}
{McBreen}, S., {McBreen}, B., {Quilligan}, F., \& {Hanlon}, L. 2002a, A\&A, 385,
  L19

\bibitem[{{McBreen} {et~al.}(2002b){McBreen}, {McBreen}, {Hanlon},
  \& {Quilligan}}]{mcbreenb:2002}
{McBreen}, S., {McBreen}, B., {Hanlon}, L., \& {Quilligan}, F.
  2002b, (A\&AL in press)

\bibitem[{{McBreen} {et~al.}(2001){McBreen}, {Quilligan}, {McBreen}, {Hanlon},
  \& {Watson}}]{sheila:2001}
{McBreen}, S., {Quilligan}, F., {McBreen}, B., {Hanlon}, L. \& {Watson}, D.
  2001, A\&A, 380, L31

\bibitem[{{Narayan} {et~al.}(2001){Narayan}, {Piran}, \&
  {Kumar}}]{narayan:2001}
{Narayan}, R., {Piran}, T. \& {Kumar}, P. 2001, ApJ, 557, 949

\bibitem[{Nakar \& Piran(2002)}]{nakar:2002}
Nakar, E. \& Piran, T. 2002, ApJ, 572, L139

\bibitem[{{Norris}(2002)}]{norris:2002}
{Norris}, J.~P. 2002, [astro-ph/0201503]

\bibitem[{{Paczynski}(1998)}]{pacy:1998}
{Paczynski}, B. 1998, ApJ, 494, L45

\bibitem[{Panaitescu \& Kumar(2001)}]{panait:2001}
Panaitescu, A. \& Kumar, P. 2001, ApJ, 560, L49

\bibitem[{{Palmer}(1999)}]{palmer:1999}
{Palmer}, D.~M. 1999, ApJ, 512, L113

\bibitem[{{Piran}(1999)}]{piran:1999}
{Piran}, T. 1999, Phys. Report, 314, 575

\bibitem[{{Piran} {et~al.}(2001)}]{piran:2001}
{Piran}, T. {Kumar}, P. {Panaitescu}, A. \& {Piro}, L. 2001, ApJ, 560, L167

\bibitem[{{Popham} {et~al.}(1999){Popham}, {Woosley}, \& {Fryer}}]{popham:1999}
{Popham}, R., {Woosley}, S.~E. \& {Fryer}, C. 1999, ApJ, 518, 356

\bibitem[{{Preece} {et~al.}(2000){Preece}, {Briggs}, {Mallozzi}, \& {et
  al}}]{pebs:2000}
{Preece}, R.~D., {Briggs}, M.~S., {Mallozzi}, R.~S. {et al}. 2000, ApJS,
  126, 19

\bibitem[{{Quilligan} {et~al.}(2002){Quilligan}, {McBreen}, {Hanlon}, {et al}}]{quilligan:2002}
{Quilligan}, F., {McBreen}, B., {Hanlon}, L. {et al}. 2002, A\&A, 385, 377 

\bibitem[{{Ramirez-Ruiz} \& {Fenimore}(2000)}]{ramirez:2000}
{Ramirez-Ruiz}, E. \& {Fenimore}, E.~E. 2000, ApJ, 539, 712

\bibitem[{{Ramirez-Ruiz} \& {Merloni}(2001)}]{rm:2001}
{Ramirez-Ruiz}, E. \& {Merloni}, A. 2001, MNRAS, 320, L25

\bibitem[{{Rees} \& {M{\' e}sz{\' a}ros}(1994)}]{reemes:1994}
{Rees}, M.~J. \& {M{\' e}sz{\' a}ros}, P. 1994, ApJ, 430, L93

\bibitem[{{Reeves} {et~al.}(2002){Reeves}, {Watson}, {Osborne}, {et al}}]{reeves:2002}
{Reeves}, J.~N., {Watson}, D., {Osborne}, J.~P. {et al.} 2002, Nature, 416, 512

\bibitem[{{Ruffert} \& {Janka}(1999)}]{ruffjan:1999}
{Ruffert}, M. \& {Janka}, H.~T. 1999, A\&A, 344, 573

\bibitem[{{Salmonson}(2000)}]{sal:2000}
{Salmonson}, J.~D. 2000, ApJ, 544, L115

\bibitem[{{van Paradijs} {et~al.}(1997){van Paradijs}, {Groot}, {Galama},
  {et al}}]{vanpara:1997}
{van Paradijs}, J., {Groot}, P.~J., {Galama}, T. {et al.} 1997, Nature, 386, 686

\bibitem[{{van Paradijs} {et~al.}(2000){van Paradijs}, {Kouveliotou}, \&
  {Wijers}}]{van:2000}
{van Paradijs}, J., {Kouveliotou}, C. \& {Wijers}, R.~A.~M.~J. 2000, Ann. Rev. Astron. Ap., 38,
  379
  
\bibitem[{{Vietri} \& {Stella}(1998)}]{vietri:1998}
{Vietri}, M. \& {Stella}, L. 1998, ApJ, 507, L45

\end{thebibliography}
\bibliographystyle{apj}
\end{document}